\documentclass{article}
\pdfoutput=1
\usepackage[utf8]{inputenc}
\usepackage[a4paper]{geometry}
\usepackage[backend=bibtex]{biblatex}
\bibliography{bib.bib}
\usepackage{hyperref}
\usepackage{tensor}
\usepackage{amsmath}
\usepackage{color}
\usepackage{mathtools}
\usepackage{amsfonts}
\usepackage{graphicx}
\usepackage{subfig}

\newtheorem{theorem}{Theorem}

\usepackage{newunicodechar}
\newunicodechar{∂}{\partial}
\newunicodechar{∀}{\forall}
\newunicodechar{×}{\times}
\newunicodechar{α}{\alpha}
\newunicodechar{β}{\beta}
\newunicodechar{γ}{\gamma}
\newunicodechar{δ}{\delta}
\newunicodechar{ε}{\epsilon}
\newunicodechar{η}{\eta}
\newunicodechar{φ}{\varphi}
\newunicodechar{ϕ}{\phi}
\newunicodechar{λ}{\lambda}
\newunicodechar{ψ}{\psi}
\newunicodechar{ρ}{\rho}
\newunicodechar{ω}{\omega}
\newunicodechar{τ}{\tau}
\newunicodechar{θ}{\theta}
\newunicodechar{μ}{\mu}
\newunicodechar{ν}{\nu}
\newunicodechar{π}{\pi}
\newunicodechar{σ}{\sigma}
\newunicodechar{ξ}{\xi}
\newunicodechar{Γ}{\Gamma}
\newunicodechar{Ψ}{\Psi}
\newunicodechar{Ω}{\Omega}
\newunicodechar{Σ}{\Sigma}
\newunicodechar{Θ}{\Theta}
\newunicodechar{Λ}{\Lambda}

\newcommand{\D}{\mathrm{d}}
\newcommand{\dx}{\mathrm{d}x}

\newcommand{\ds}{\mathrm{d}s}

\newcommand{\N}[1]{\mathcal{N}\indices{#1}}

\newcommand{\p}[1]{P\indices{#1}}

\newcommand{\bel}[1]{\begin{equation}\label{#1}}
\newcommand{\ee}{\end{equation}}
\newcommand{\nn}{\nonumber}

{\catcode `\@=11 \global\let\AddToReset=\@addtoreset}
\AddToReset{equation}{section}

{\catcode `\@=11 \global\let\AddToReset=\@addtoreset}
\AddToReset{figure}{section}

{\catcode `\@=11 \global\let\AddToReset=\@addtoreset}
\AddToReset{table}{section}

\title{Timelike singularities and Hamiltonian cosmological billiards \footnote{UWThPh-2015-33}}
\author{Paul Klinger\\
Faculty of Physics, University of Vienna\\
Boltzmanngasse 5, 1090 Vienna, Austria}
\begin{document}
\maketitle
\begin{abstract}
We construct a large class of vacuum solutions of the Einstein equations without any symmetries and with controlled asymptotics near a timelike singularity. The solutions are obtained by a Fuchs analysis of the equations which evolve the metric in a spacelike direction.

We further observe that the changes of signs of some of the terms (walls) in the associated Hamiltonian invalidate the  ``cosmological billards''  heuristic arguments for existence of singularities of mixmaster type in the current context.
\end{abstract}

\section{Introduction}
An important issue in general relativity is the nature of singularities. While it is widely believed that the strong cosmic censorship conjecture holds, which can be loosely stated as the expectation that timelike singularities do not form by evolution from generic spatially compact or asymptotically flat initial data sets \cite{Penrose1979}, the issue is wide open. From this perspective it is of interest to consider timelike singularities and therefore the ways in which cosmic censorship could be violated.

There are numerous exact solutions with timelike singularities (e.g. \cite{Griffiths2009}). Such solutions are typically obtained in searches of solutions with symmetries. This leads naturally to the question, whether there exist solutions with timelike singularities and without symmetries. We prove in this work that this is indeed the case: We construct an analog to the class of the non-chaotic solutions without symmetries and with controlled asymptotics of \cite{Chrusciel2015,Klinger2015}, by changing the time parameter $\tau$ from a timelike to a spacelike coordinate. As the behavior of the Hamiltonian differs from the spacelike case only by sign changes which do not affect the analysis in the analytic case, we obtain a family of solutions with the same free functions and asymptotics (in terms of the now spacelike $τ$ coordinate) but with a timelike instead of spacelike singularity.

The construction of the solutions is based on the cosmological billiard formalism using the  Iwasawa decomposition of the metric. This method was introduced by Damour, Henneaux, and Nicolai in \cite{Damour2003} to give a heuristic argument for the chaotic picture of spacelike singularities provided by the  BKL conjecture, and later by Damour and DeBuyl in \cite{Damour2008} to provide a precise statement of the conjecture.

We also show that the change of the time parameter $\tau$ from a timelike to a spacelike coordinate, i.e. considering timelike instead of spacelike singularities, switches the signs of some of the terms (walls) in the Hamiltonian considered. These changes violate the property of the spacelike case that the coefficients of the dominant wall terms are positive, thus rendering the arguments of Damour et al. in \cite{Damour2003} inapplicable. The affected terms become attractive rather than repulsive, allowing subdominant walls lying behind the dominant ones to become relevant. This does not affect the class of solutions we construct here, as these are non-generic and use an ansatz that suppresses the wall terms asymptotically.

\section{Derivation of Hamiltonian for spacelike ``time''-variable}
We follow the derivation of the Hamiltonian formalism by Wald \cite[Appendix E.2]{Wald1984}. The spacetime metric is denoted by $\bar{g}_{αβ}$ while the induced lorentzian metric on the timelike hypersurfaces of constant $τ$ is denoted by $g_{ij}$.  We choose a zero shift gauge, i.e. the metric takes the form
\begin{equation}
\ds^2=N^2\Dτ^2+g_{ij}\dx^i\dx^j\,.
\end{equation}

As the hypersurfaces of constant $τ$ are timelike, their normal vector is spacelike. This means that the Gauss equation takes the form
\begin{equation}
R\indices{_{abc}^d}=g_a{}^f g_b{}^g g_c{}^k g^d{}_j\bar{R}\indices{_{fgk}^j}+K_{ac}K\indices{_b^d}-K_{bc}K\indices{_a^d}
\end{equation}
where $R\indices{_{abc}^d}$ and $\bar{R}\indices{_{abc}^d}$ are the Riemann curvature tensors of the induced and full metric respectively and $K_{ab}$ is the second fundamental form of the hypersurface. Compared to the case of spacelike hypersurfaces the signs of the $KK$ terms are interchanged.

Using $g_{αβ}=\bar{g}_{αβ}-n_α n_β$ with $n^α$ the unit normal vector of the hypersurface ($n_αn^α=1$) gives
\begin{equation}
\bar{R}_{αβγδ}g^{αγ}g^{βδ}=-2\bar{G}_{αβ}n^αn^β\,.
\end{equation}

This leads to a change of sign in the constraint equation:
\begin{equation}\label{const}
0=\bar{G}_{μν}n^μ n^ν = -\frac{1}{2}R-(K\indices{^\mu_\mu})^2+K_{μν}K^{μν}\,.
\end{equation}

Contracting the Einstein tensor twice with the normal vector $n^a$ gives an expression for the scalar curvature:
\begin{equation}\label{scurv}
\bar{R}=-2n^αn^β(\bar{G}_{αβ}-\bar{R}_{αβ})\,.
\end{equation}

The definition of the Riemann tensor gives for the last term
\begin{equation}\label{riemnn}
\begin{split}
\bar{R}_{αβ}n^αn^β=&\bar{R}_{αγβ}{}^γn^αn^β=-n^α(\nabla_α\nabla_γ-\nabla_γ\nabla_α)n^γ\\
=&(\nabla_α n^α)(\nabla_γ n^γ)-(\nabla_γ n^α)(\nabla_α n^γ)\\
&-\nabla_α(n^α\nabla_γ n^γ)+\nabla_γ(n^α\nabla_αn^γ)\\
=&(K^α_α)^2-K_{αγ}K^{αγ}-\nabla_α(n^α\nabla_γ n^γ)+\nabla_γ(n^α\nabla_αn^γ)
\end{split}
\end{equation}
where the last two terms are divergences, which will be discarded in the Lagrangian.

Using \eqref{const}, \eqref{scurv}, \eqref{riemnn} and $\sqrt{-\bar{g}}=N\sqrt{-g}$ to express the Einstein-Hilbert action gives
\begin{equation}
\mathcal{L}=\sqrt{-\bar{g}}\bar{R}=-\sqrt{-g}N\left(R-K_{ab}K^{ab}+(K^a{}_a)^2\right)\,.
\end{equation}
The momenta canonically conjugate to the components $g_{ij}$ are given by
\begin{equation}
π^{ij}=\frac{∂\mathcal{L}}{∂\dot{g}_{ij}}=\sqrt{-g}(K^{ij}-K^k{}_k g^{ij})=N^{-1}\sqrt{-g}\frac{1}{2}(\dot{g}_{kl}g^{ki}g^{lj}-\dot{g}_{kl} g^{kl}g^{ij})\,,
\end{equation}
unchanged from the standard case.

The Hamitonian, expressed in terms of the canonical coordinates $g_{ab}$ and momenta $π^{ab}$ is finally
\begin{equation}\label{hamiltonian}
\mathcal{H}=π^{ab}\dot{g}_{ab}-\mathcal{L}=(-g)^{-1/2}N\left(π^{ab}π_{ab}-\frac{1}{2}(π^a{}_a)^2\right)+RN\sqrt{-g}
\end{equation}
i.e. the standard one with the sign of the curvature term changed.

\section{Iwasawa variable Hamiltonian}
Here we will describe the changes to the derivation of the Iwasawa variable Hamiltonian, as given in Appendix A of \cite{Klinger2015}.

Since the level sets of $\tau$ are timelike, we need to decide which frame vector is the timelike one. As the Iwasawa ansatz breaks the symmetry between the frame vectors, different choices will lead to different dynamical systems. We will use an index  $J\in\{1,2,3\}$ to distinguish between those cases: $x^J$ will denote the timelike coordinate.

The Lorentzian metric $g_{ij}$ on the $τ=\text{const}$ hypersurfaces is split in Iwasawa variables as
\begin{equation}
g_{ij}=\sum_a m^J_a e^{-2β^a}\N{^a_i}\N{^a_j}
\end{equation}
where $m^J_a=1-2δ_{Ja}$, i.e. $-1$ for $a=J$ and $1$ otherwise. 

We set the lapse function $N$ equal to $\sqrt{-g}$ where $g$ is the determinant of the metric $g_{ij}$. The (timelike) singularity will be approached as $τ\to \infty$.

The conjugate momenta $π_a$to the $β^a$ and $P\indices{^i_a}$ to the $\N{^a_i}$ are given by
\begin{equation}
π_a=\frac{∂\mathcal L}{∂\dot{β}^a}=\frac{∂\mathcal L}{∂\dot{g}_{ij}}\frac{∂\dot{g}_{ij}}{∂\dot{β^a}}=-2π^{ij}m^J_ae^{-2β^a}\N{^a_i}\N{^a_j}
\end{equation}
and
\begin{equation}
P\indices{^i_a}=2m^J_a π^{ij}\N{^a_j}e^{-2β^a}\,,
\end{equation}
i.e. the same as in the spacelike case except for the additional factor $m^J_a$.

The non-curvature terms of the Hamiltonian \eqref{hamiltonian}, with $N=\sqrt{-g}$ inserted, are
\begin{equation}\label{2015X30.2}
π^{ab}π_{ab}-\frac{1}{2}(π^a{}_a)^2\,.
\end{equation}

The first term can be split into
\begin{equation}\label{2015X30.1}
\frac{1}{4}\sum_a π_a^2+\frac{1}{2}\sum_{a<b}m^J_a m^J_b e^{-2(β^b-β^a)}\left(P\indices{^j_a}\N{^b_j}\right)^2\,.
\end{equation}
The first term of \eqref{2015X30.1} together with the second term of \eqref{2015X30.2} give the kinetic term $G^{ab}π_aπ_b$ of the Hamiltonian, unchanged from the spacelike case. The second term of \eqref{2015X30.1} is the symmetry wall term, with the addition of $m^J_a$ and $m^J_b$. These cause a sign change for two of the symmetry walls.

The Iwasawa form of the curvature term in the Hamiltonian \eqref{hamiltonian}, which gives the gravitational potential walls, is calculated in the Iwasawa frame, where the metric takes the form
\begin{equation}
γ_{ab}=δ_{ab}m^J_ae^{-2β^a}=δ_{ab}m^J_a A_a^2
\end{equation}
with $A_a^2:=\exp(-2β^a)$ i.e. with an additional factor $m^J_a$ compared to the spacelike case.

This is the only change in the derivation of the curvature term, as the Cartan formulas remain unchanged. The terms corresponding to the dominant gravitational walls are
\begin{equation}
\frac{1}{4}\hspace{-0.9em}\sum_{a\neq b\neq c \neq a}\hspace{-0.9em}(C\indices{^a_b_c})^2\frac{A_a^2}{A_c^2A_b^2}\frac{m^J_a}{m^J_b m^J_c}\,.
\end{equation}
In $3+1$ dimensions exactly one of $a, b, c$ is equal to $J$, which adds an additional minus in front of this term. This cancels the change of sign in the Hamiltonian.

In addition there are sign changes in the subdominant gravitational terms but as they have indeterminate sign even for the spacelike case this does not affect the analysis.

In conclusion, for $3+1$ dimensions, the prefactors of two of the symmetry walls change sign. In the case $J=1$ this involves the $β^2-β^1$ and $β^3-β^1$ walls, for $J=2$ the $β^3-β^2$ and $β^2-β^1$ walls and for $J=3$ the $β^3-β^2$ and $β^3-β^1$ walls. In all cases the sign of at least one dominant wall term changes.
 
\begin{figure}[p]

    \centering
    \subfloat[$τ$ timelike, $β^3=3$ slice]{
        \includegraphics[width=0.45\textwidth]{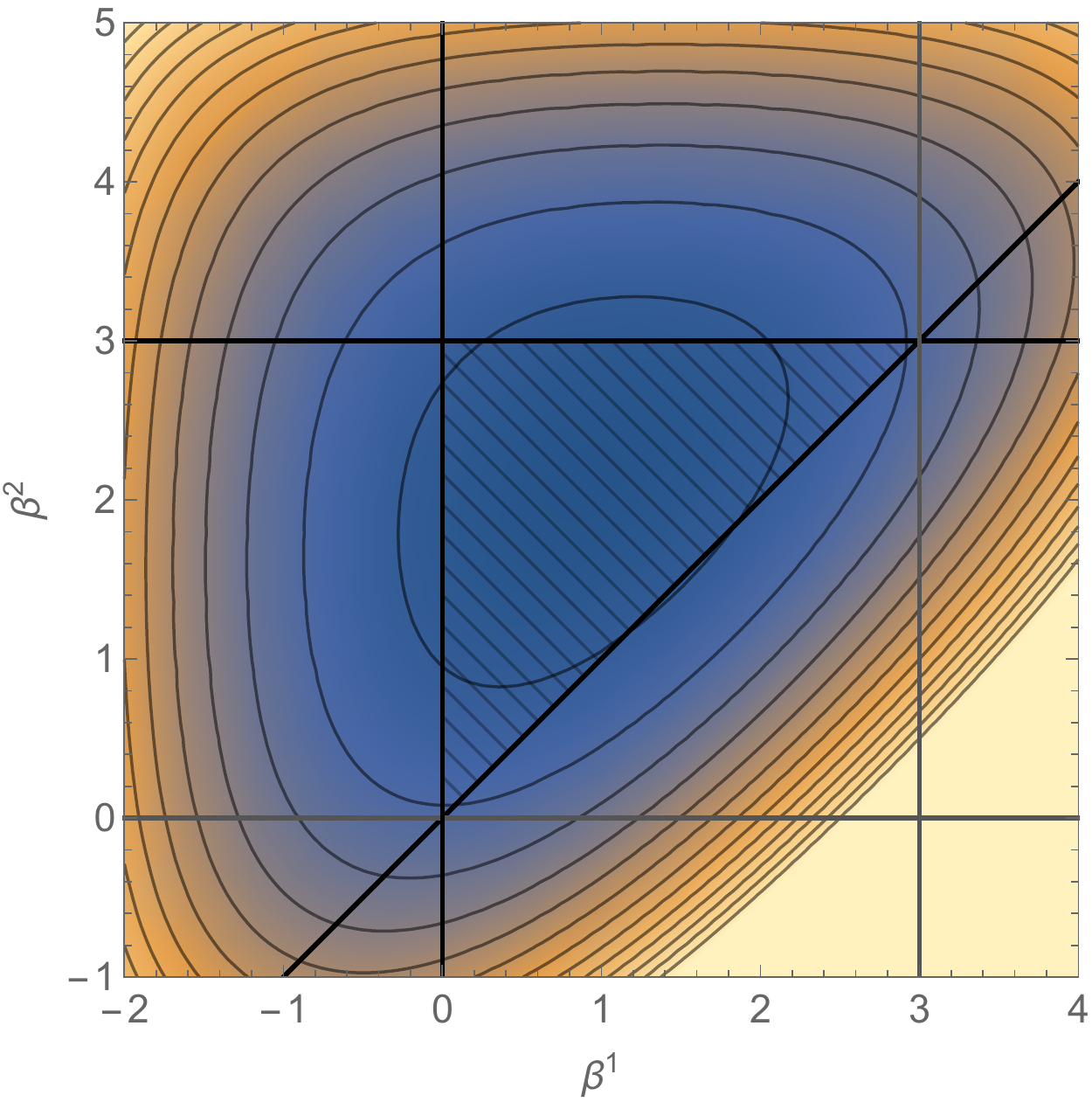}
	} 
    \subfloat[$τ$ timelike, $β^1=1$ slice]{
        \includegraphics[width=0.45\textwidth]{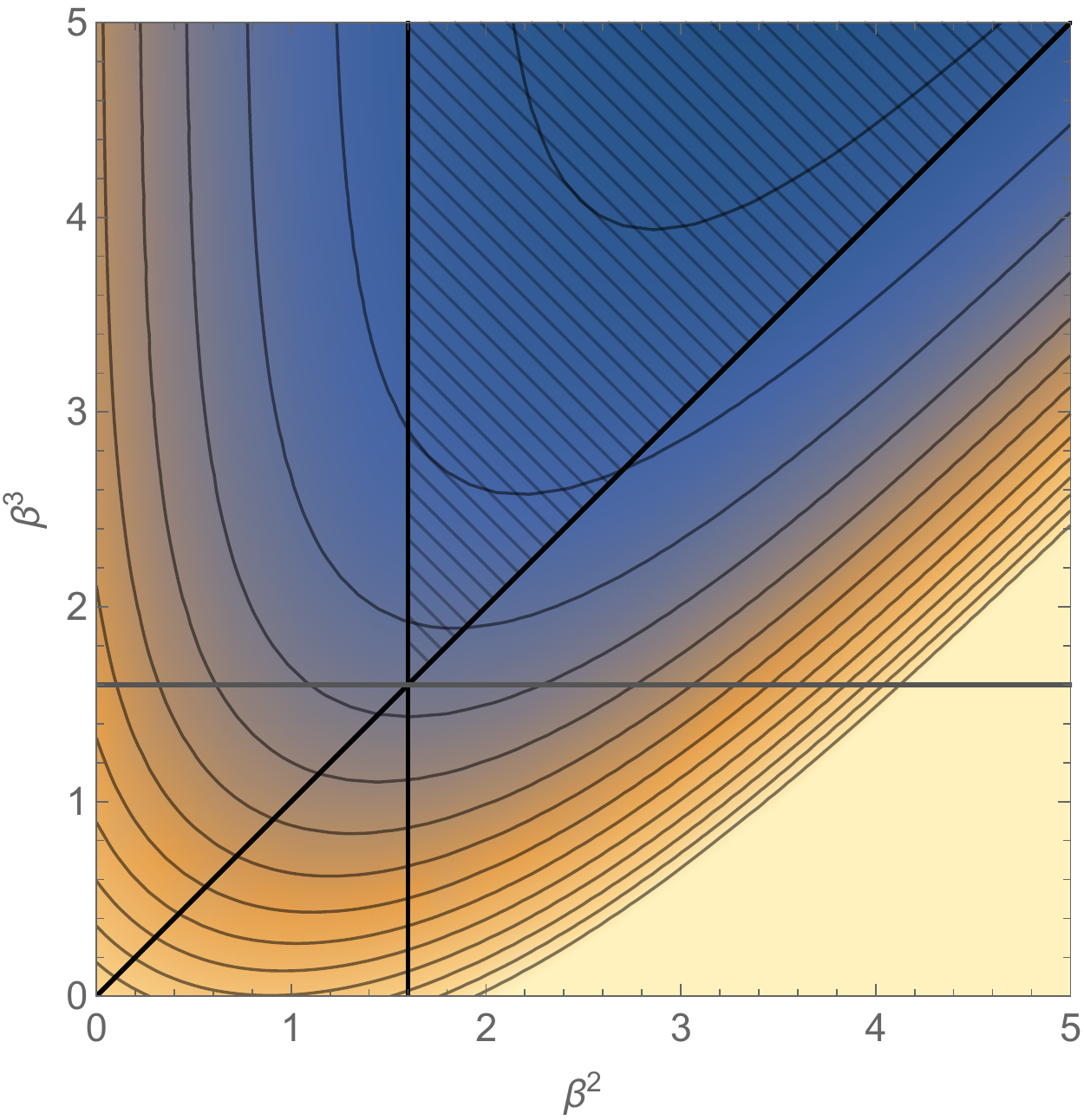}
    }
    \caption{Sketch of the potential for the case of timelike $τ$ (i.e. with spacelike singularity). Only the exponential terms are plotted, the coefficients are set to $1$. The potential increases from dark blue through orange to light yellow. Black lines mark the dominant walls, grey lines the subdominant walls and the allowed region (i.e. the ``billiard table'') is hatched.}
	\label{fig:t}
\end{figure}

\begin{figure}[p]

	\centering
    \subfloat[$x^1$ timelike, $β^3=3$ slice]{
        \includegraphics[width=0.45\textwidth]{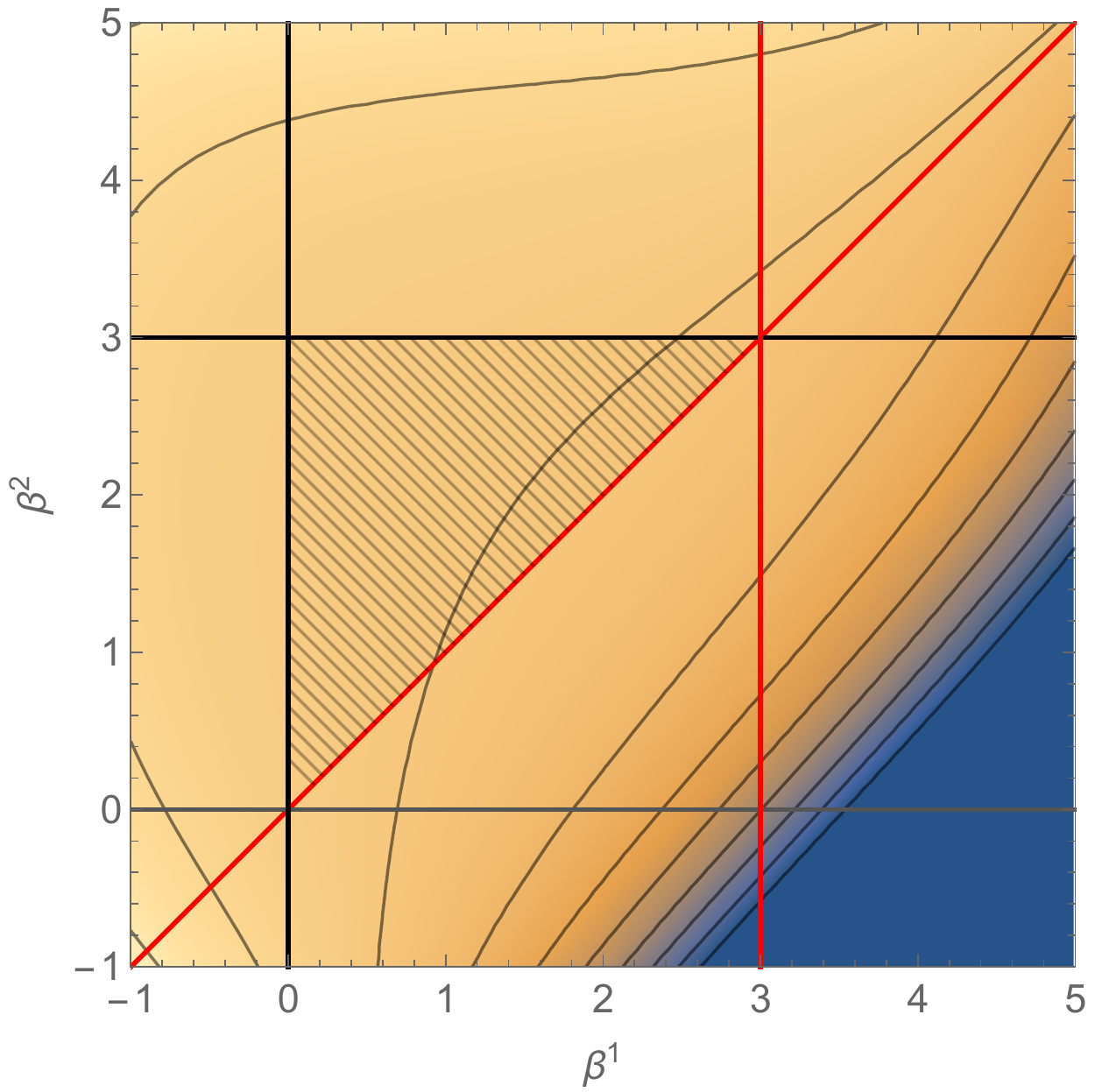}
	}
    \subfloat[$x^1$ timelike, $β^1=1$ slice]{
        \includegraphics[width=0.45\textwidth]{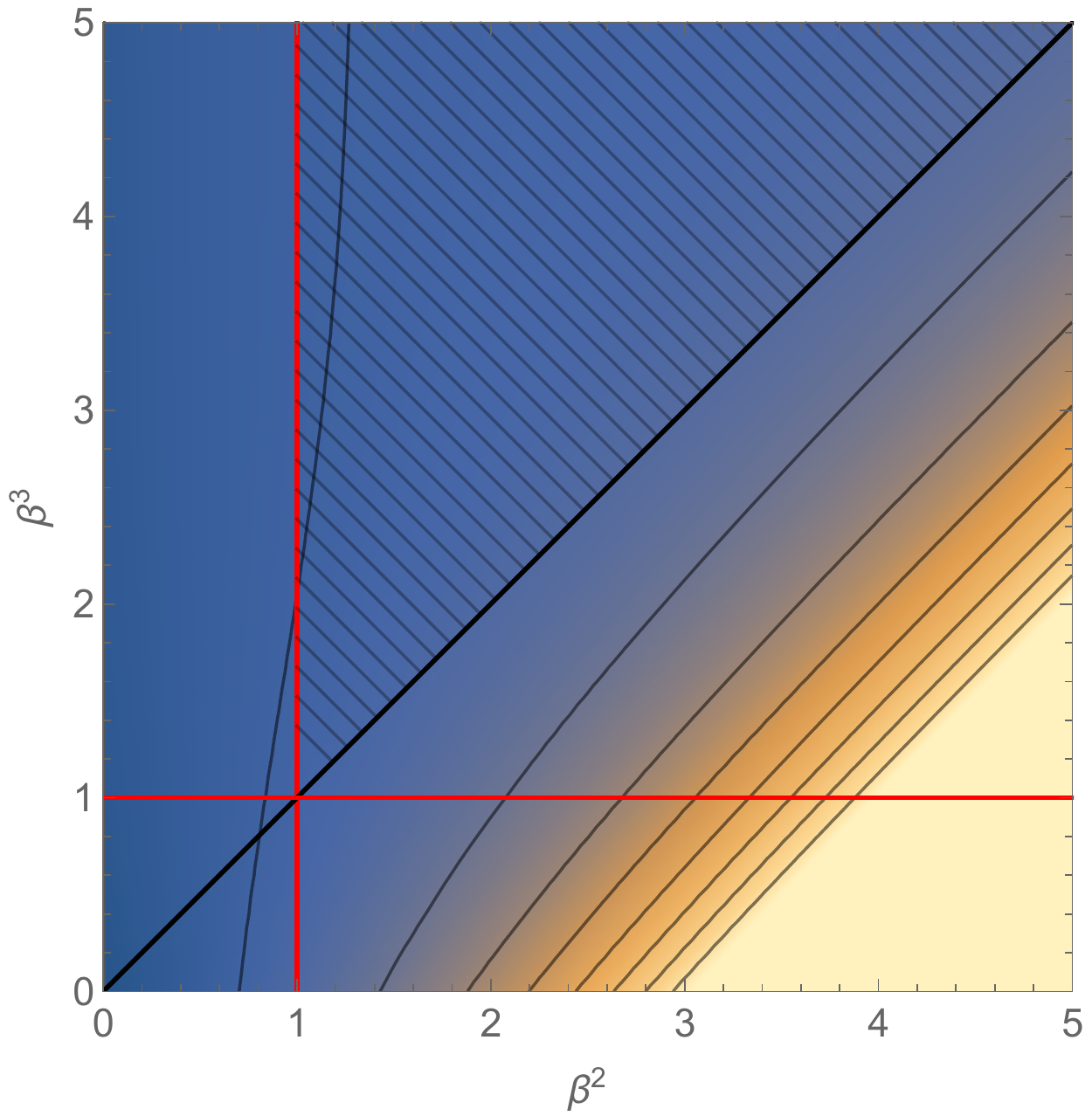}
    }
    \caption{$J=1$, i.e. $x^1$ timelike case. The walls with negative coefficients (set to $-1$ in the plots) are marked by red lines. The hatched region is the same as in figure \ref{fig:t} but no longer corresponds to an allowed region, as the potential approaches $-\infty$ outside it.}
    \label{fig:x1}
\end{figure}

\begin{figure}[p]
	\centering
    \subfloat[$x^2$ timelike, $β^3=3$ slice]{
        \includegraphics[width=0.45\textwidth]{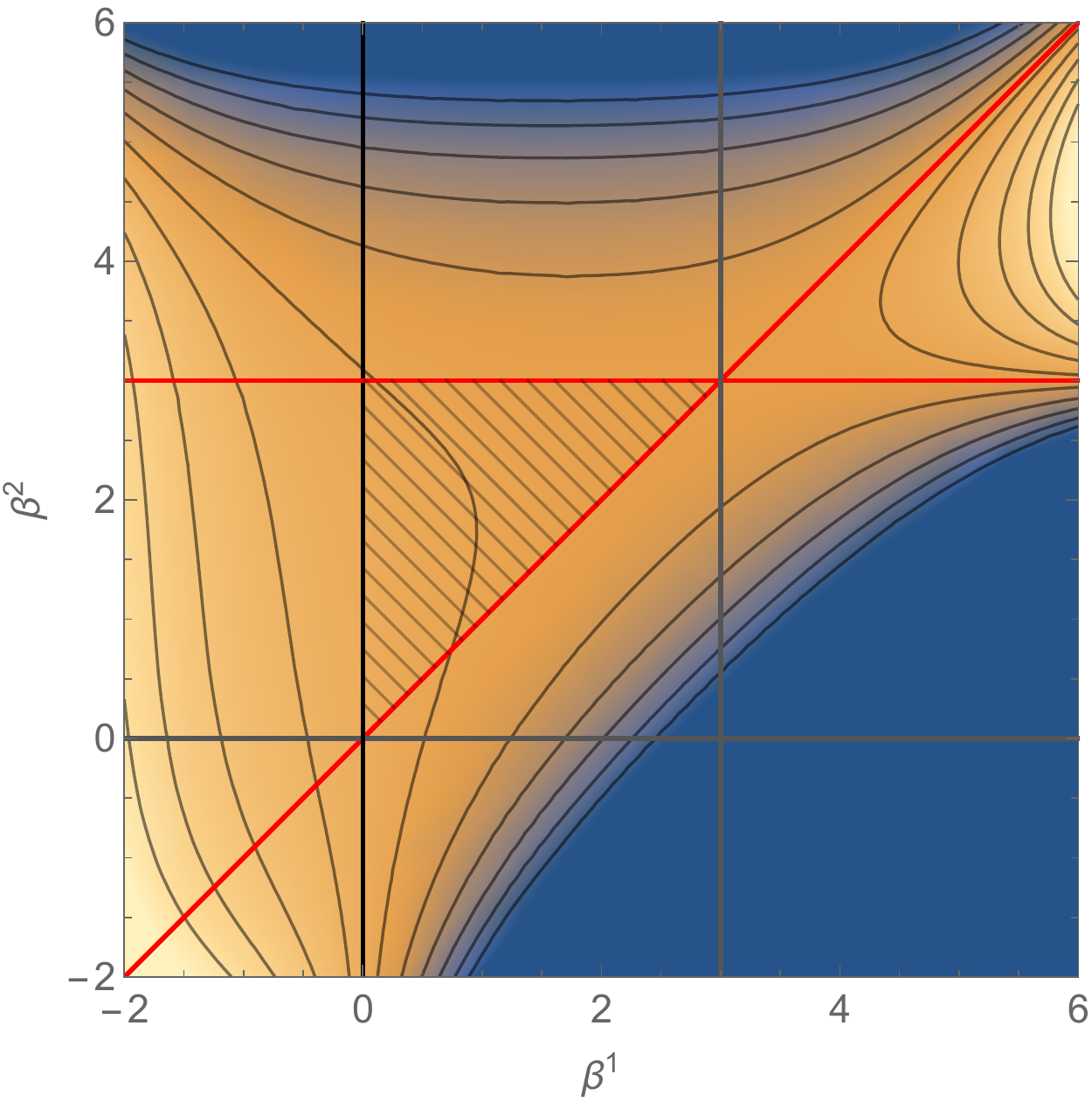}
	}
    \subfloat[$x^2$ timelike, $β^1=2$ slice]{
        \includegraphics[width=0.45\textwidth]{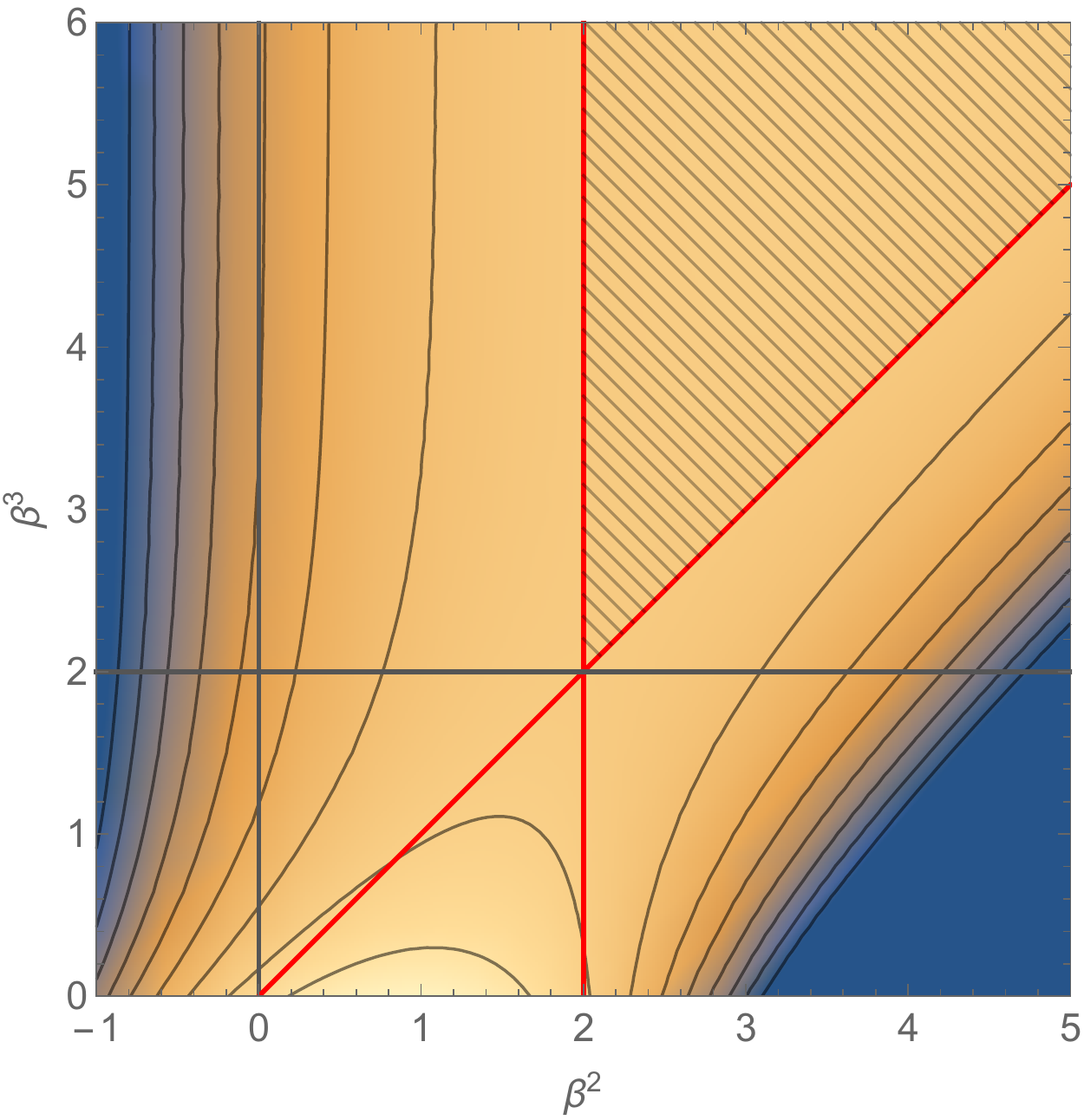}
    }
    \caption{$J=2$ i.e. $x^2$ timelike case.}
    \label{fig:x2}
\end{figure}

\begin{figure}[p]

	\centering
    \subfloat[$x^3$ timelike, $β^3=3$ slice]{
        \includegraphics[width=0.45\textwidth]{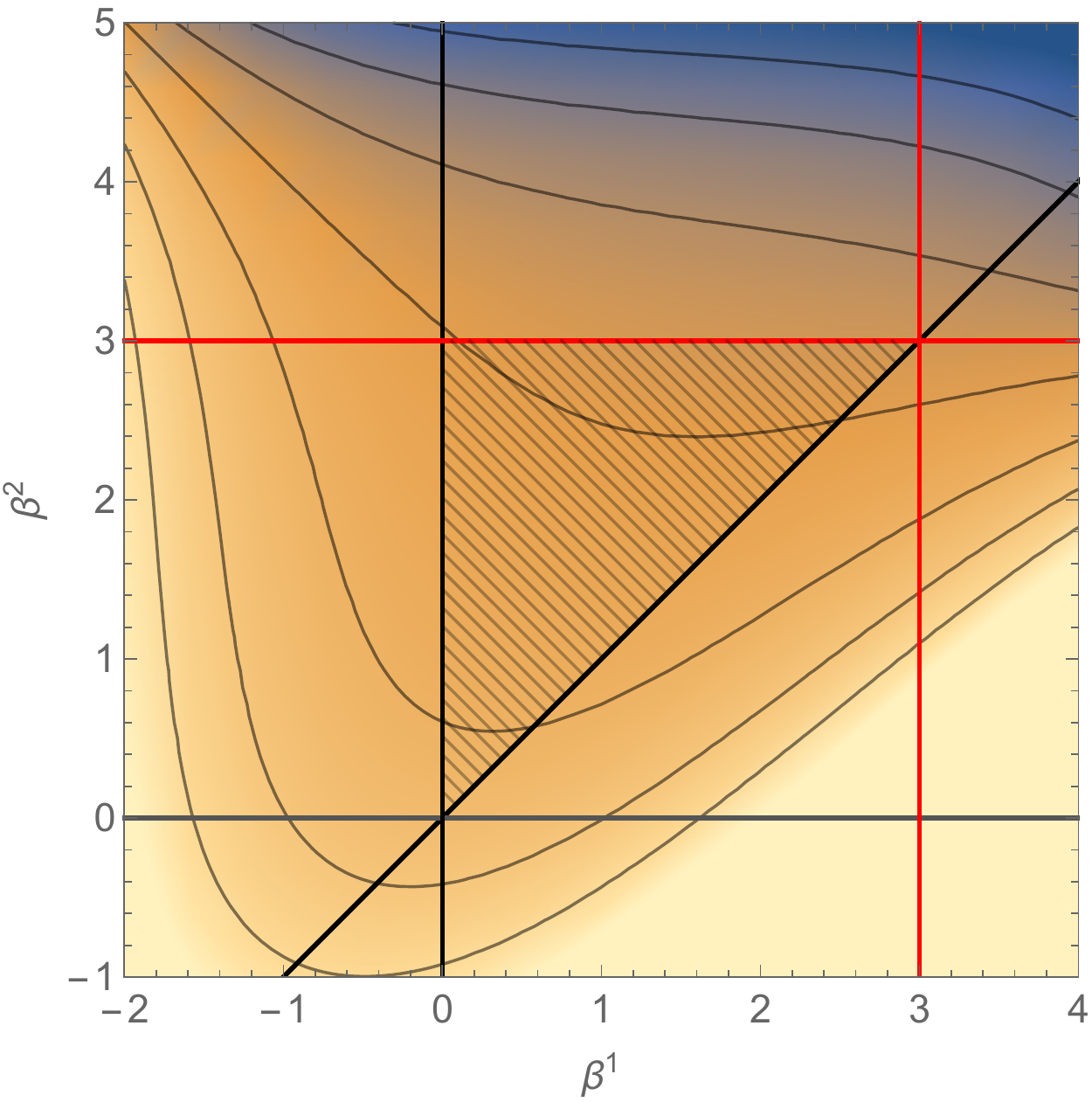}
	} 
    \subfloat[$x^3$ timelike, $β^1=2$ slice]{
        \includegraphics[width=0.45\textwidth]{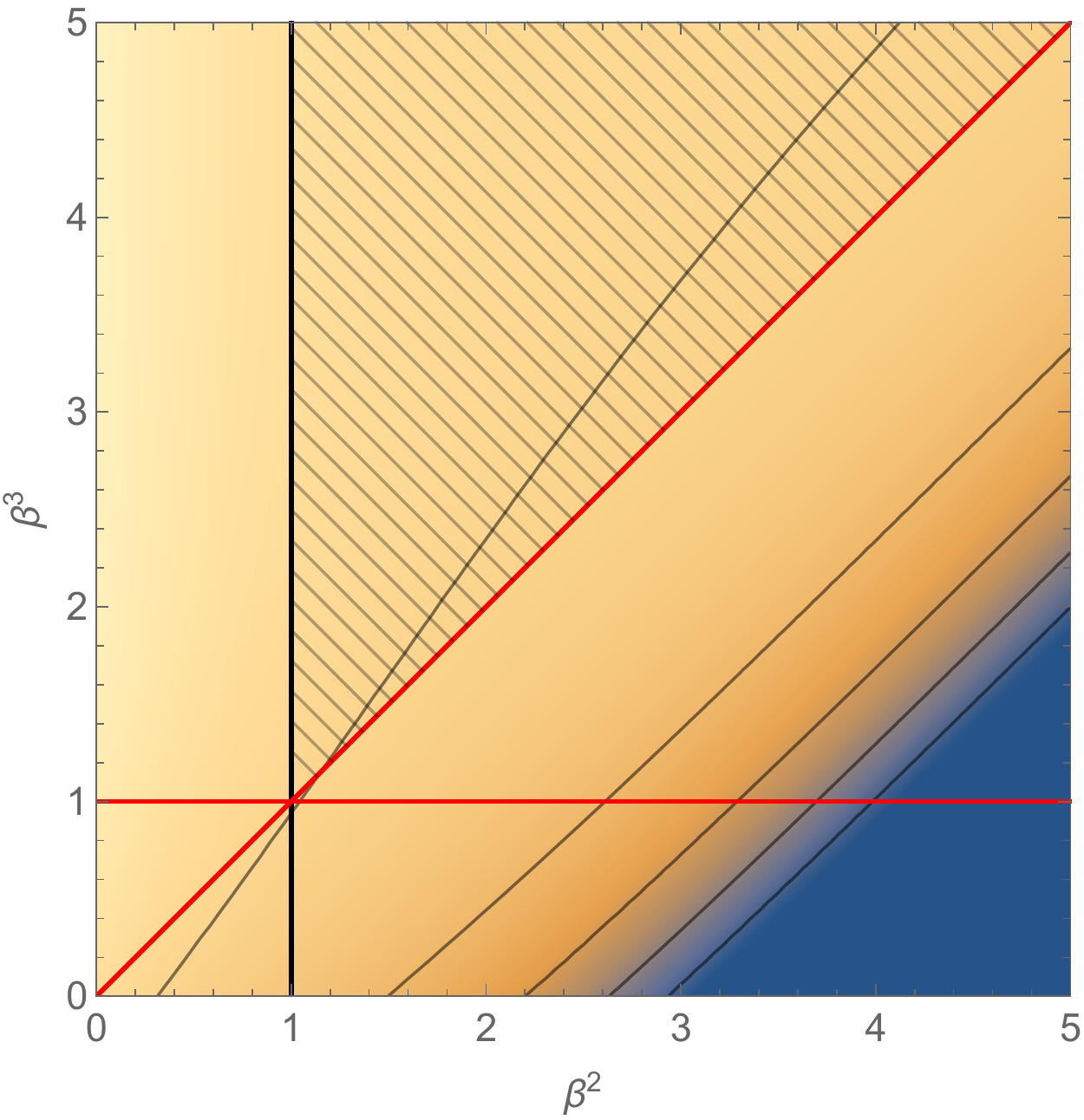}
    }
    \caption{$J=3$ i.e. $x^3$ timelike case.}
    \label{fig:x3}
\end{figure}

 The potential (i.e. the Hamiltonian without the terms containing $π^a$) for the spacelike case and the 3 choices of $J$ is sketched in figures \ref{fig:t} to \ref{fig:x3}.
 
\begin{figure}[htbp]
	\centering
    \subfloat[$τ$ timelike (spacelike singularity)]{
        \includegraphics[width=0.44\textwidth]{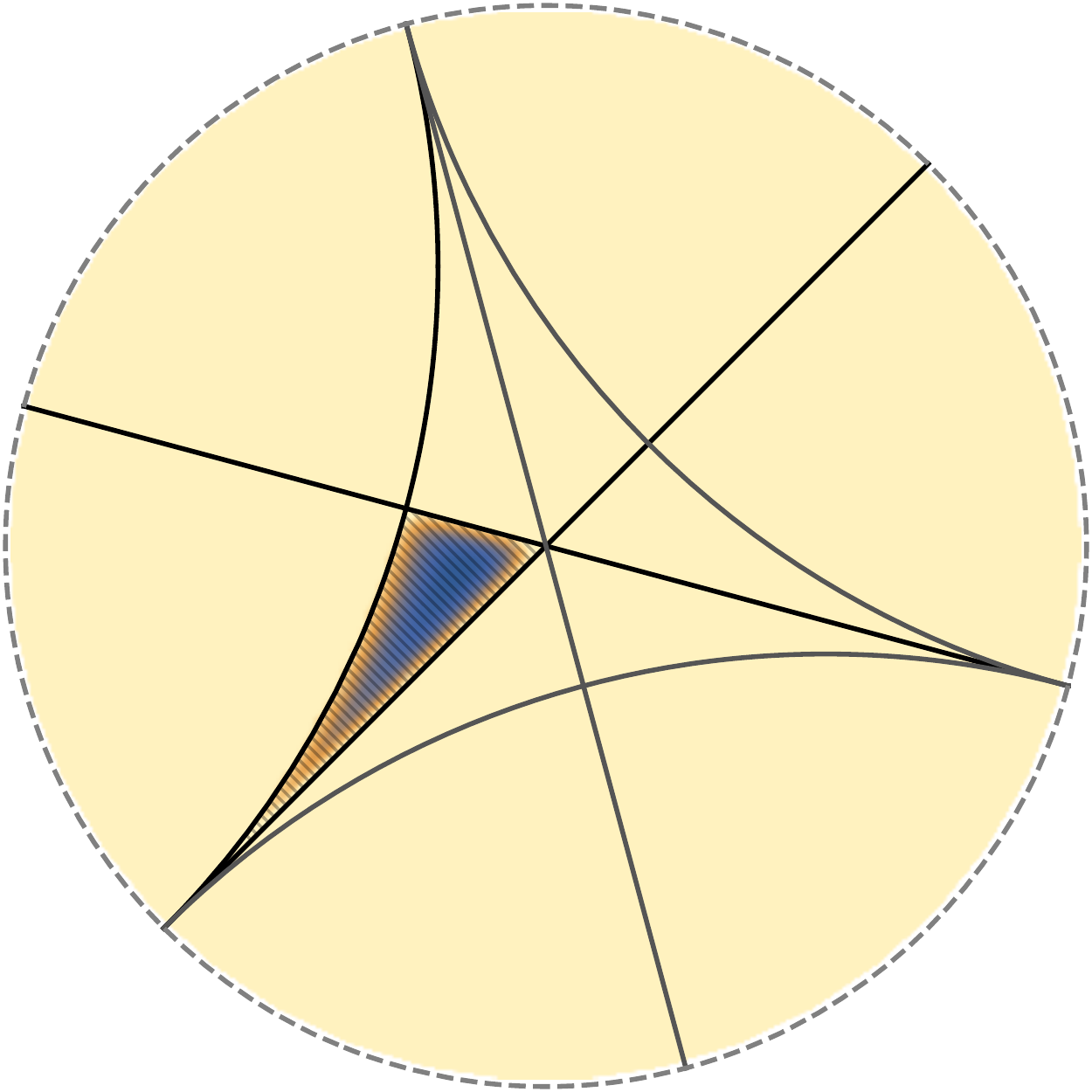}
	} 
    \subfloat[$J=1$, i.e. $x^1$ timelike]{
        \includegraphics[width=0.44\textwidth]{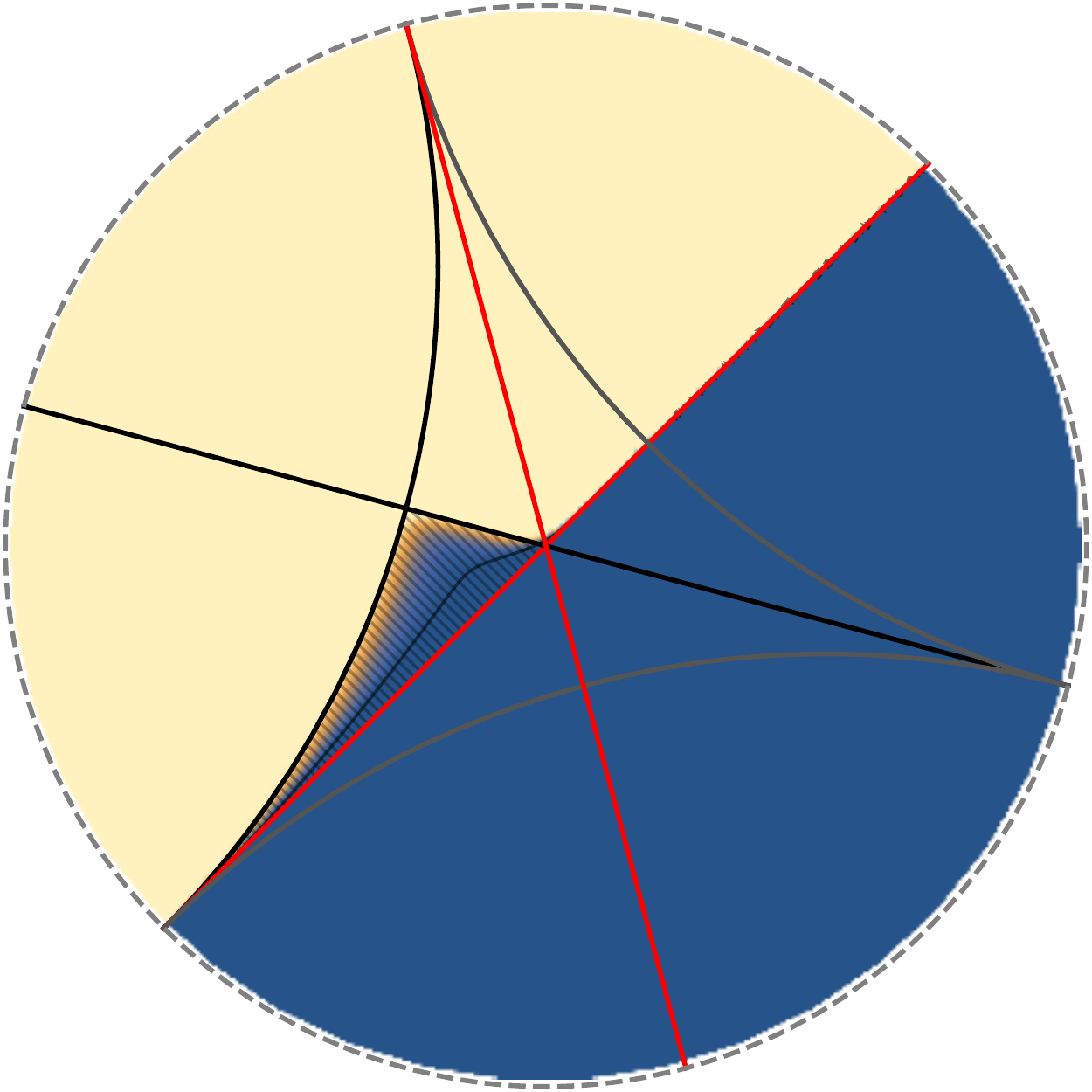}
    }\\
    \subfloat[$J=2$, i.e. $x^2$ timelike]{
        \includegraphics[width=0.44\textwidth]{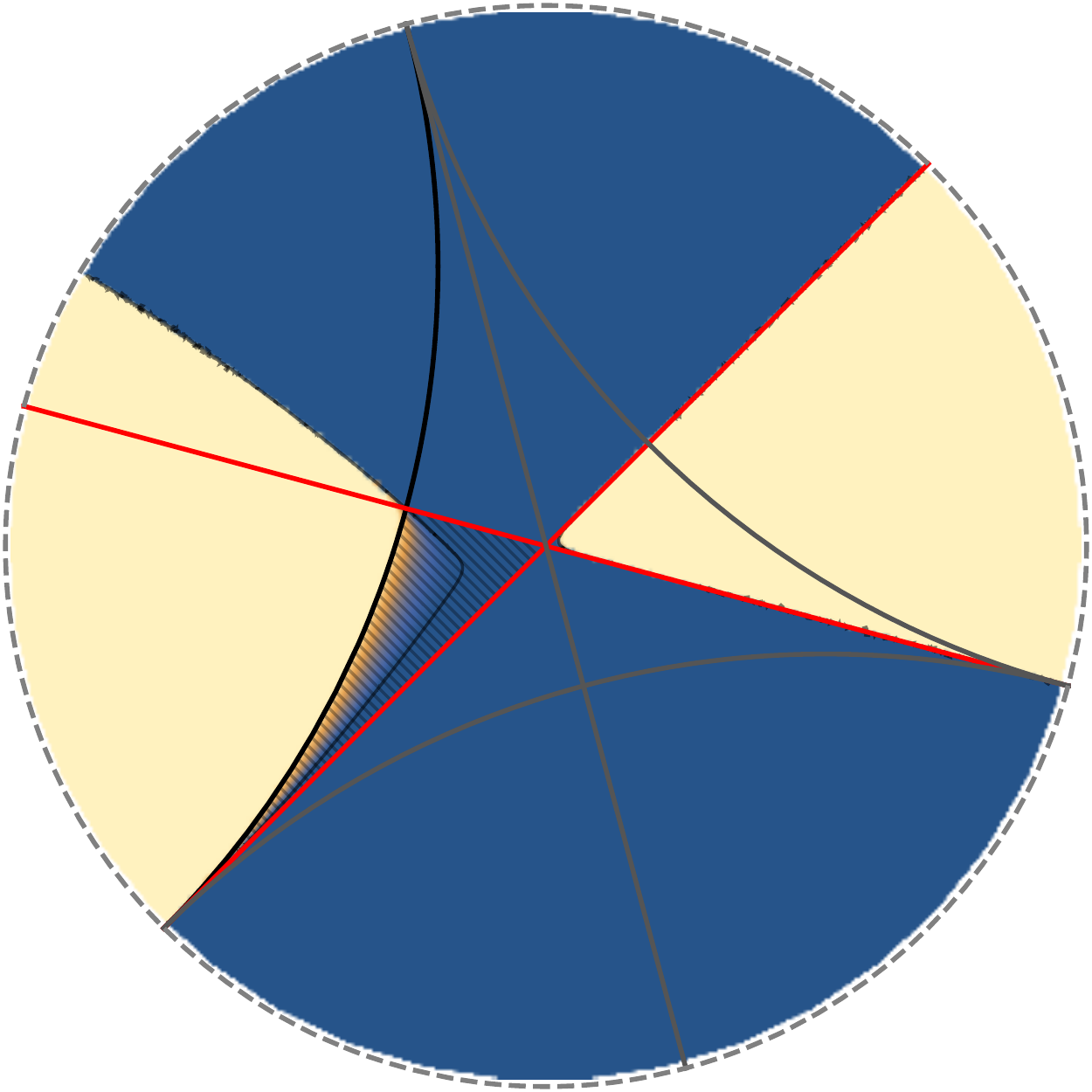}
	} 
    \subfloat[$J=3$, i.e. $x^3$ timelike]{
        \includegraphics[width=0.44\textwidth]{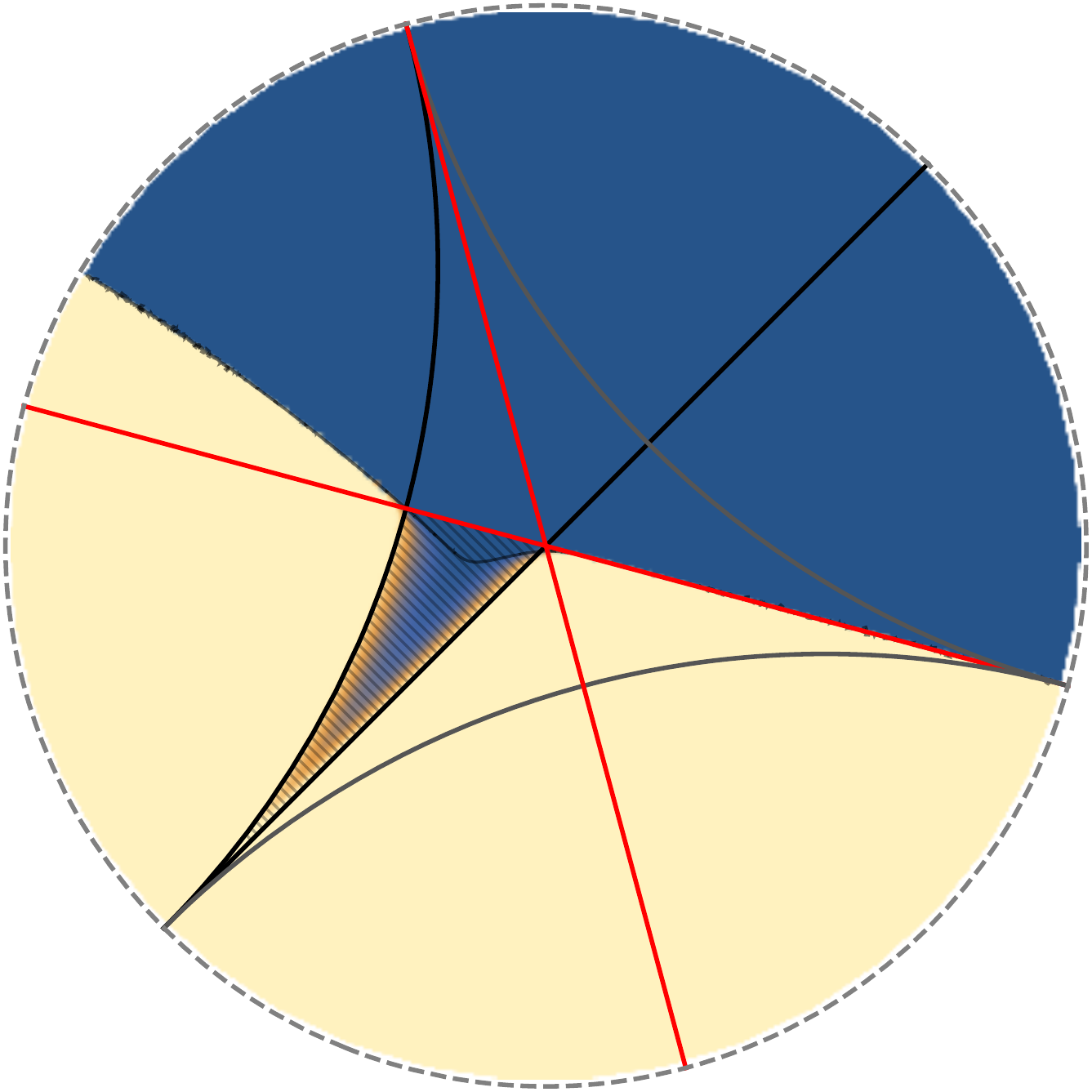}
    }
    \caption{Sketch of the potentials in the hyperbolic space of the $γ^a$, projected onto the Poincar\'{e} disk. As before the allowed region for the case of spacelike singularity is hatched, thick black lines mark the dominant walls, thick gray lines the subdominant walls and red lines the walls with negative coefficients.}
    \label{fig:hyp}
\end{figure}

\section{Consequences for cosmological billiard}
The arguments for the asymptotic billiard picture depend on the positive sign of the prefactors of the dominant wall terms: In hyperbolic coordinates $ρ$ and $γ^a$, such that $G_{ab}γ^aγ^b=-1$ ($G_{ab}$ is the constant matrix $G_{ab}=-\sum_{c\neq d}δ^c_aδ^d_b$) and $β^a=ρ γ^a$ (such a splitting is possible, assuming the solution is close to a Kasner state asymptotically \cite{Damour2003}) the Hamiltonian takes the form
\begin{equation}
H=\frac{1}{4}(-ρ^2π_ρ^2+π_γ^2)+ρ^2\sum_Ac_Ae^{-2ρ ω_A(γ)}\,.
\end{equation}
If the prefactors $c_A$ for the dominant walls are positive this approaches the ``sharp wall Hamiltonian''
\begin{equation}
H=\frac{1}{4}(-ρ^2π_ρ^2+π_γ^2)+\sum_{A'}Θ_\infty(-ω_{A'}(γ))
\end{equation}
where the sum over $A'$ only covers the dominant walls and
\begin{equation}
Θ_\infty(x)=
\begin{cases}
  0&x<0\,,\\
  \infty&x>0\,.
\end{cases}
\end{equation}
If, however, some of the prefactors are negative the corresponding terms are potential wells instead of walls. In the timelike case in 3+1 dimensions this affects at least one of the dominant symmetry walls.

Figure \ref{fig:hyp} shows the potentials in the hyperbolic space of the $γ^a$, projected onto the Poincar\'{e} disk.

\section{Consequences for solutions constructed in  \cite{Chrusciel2015,Klinger2015}}
The class of solutions constructed in \cite{Chrusciel2015,Klinger2015} for the case of timelike $τ$ also exists for spacelike $τ$. The sign changes in the Hamiltonian have no effect on the arguments concerning the evolution equations in the context of the analytic Fuchs theorem, as the decay of the exponential terms is unchanged.

An additional factor $m^J_b$ appears in the term $\tilde{π}^b{}_a$, which enters  in the Iwasawa variable momentum constraint:
\begin{equation}\label{pi_iwa}
	\tilde{π}\indices{^b_a}=\frac{1}{2}\begin{cases}
	-π_b&\text{for }a=b\,,\\
	\N{^b_i}P\indices{^i_a}&\text{for }b>a\,,\\
	m^J_be^{-2(β^a-β^b)}\N{^a_i}P\indices{^i_b}&\text{for }a>b\,.
	\end{cases}
\end{equation}
As the factor $m^J_b$ is only present in the asymptotically decaying case $a>b$, which is discarded in the asymptotic constraints, this leaves the conditions on the free functions unchanged.

Similarly there are sign changes in the derivation of the evolution equations for the constraints, given in Appendix D of \cite{Klinger2015}, which cancel out in the final equations.

As in the case of a spacelike singularity, the presence of a cosmological constant does not affect the result (see Appendix F of \cite{Klinger2015}).

This leads to the following theorem, in close analogy with the results of \cite{Chrusciel2015,Klinger2015}:

\begin{theorem}
For any choice of $J\in\{1,2,3\}$ and analytic functions $β_\circ^2$, $β_\circ^3$ and $P\indices{_\circ^2_1}$
depending on coordinates $x^i\,,i\in\{1,2,3\}$,
and for any two analytic functions, $p_\circ^2$ and $p_\circ^3$ depending on $x^i$, which satisfy the inequalities
\bel{13VI15.1}
0<p_\circ^2<(\sqrt{2}-1)p_\circ^3\,.
\ee
we obtain a solution of the vacuum Einstein equations with arbitrary cosmological constant given by the metric
\begin{equation}\label{10II15.1}
g=e^{-2\sum_{a=1}^3  β^a}\D τ^2+\sum_{a=1}^3m^J_a  e^{-2β^a}\N{^a_i}\N{^a_j}\dx^i\dx^j\,.
\end{equation}
Here $m^J_a=1-2δ_{Ja}$, $β^a$ and $\N{^a_i}$, $i,a\in\{1,2,3\}$,  depend on all coordinates $τ$, $x^i$ and behave asymptotically as
\begin{equation}\label{10II15.2}
β^a=β_\circ^a+τp_\circ^a+O(e^{-τν})\quad\text{and}\quad
\N{^a_i}=:δ^a_i+ \N{_s^a_i}=δ^a_i+O(e^{-τν})
 \,,
\end{equation}
where $ν$ is a positive constant, the $β_\circ^a$'s and  $p_\circ^a$'s depend only upon $x^i$ and $\N{_s^a_i}=0$ for $a\ge i$ with the non-vanishing terms given by
\begin{align}\label{16VI15.4}
    \N{_s^1_2}
     &=
     -\frac{\p{_\circ^2_1}e^{-2(β_\circ^2-\beta_\circ^1)}}{2(p_\circ^2-p_\circ^1)}
      e^{-τ(2p_\circ^2-2p_\circ^1 )}+O(e^{-τ(2p_\circ^2-2p_\circ^1 +ν)})
     \,,
\\
 \label{16VI15.5}
    \N{_s^2_3}
     & =
       -\frac{\p{_\circ^3_2}e^{-2(β_\circ^3-β_\circ^2)}}{2(p_\circ^3-p_\circ^2)}
         e^{-τ(2p_\circ^3-2p_\circ^2 )}
      +O(e^{-τ(2p_\circ^3-2p_\circ^2 +ν)})
      \,,
\\
    \N{_s^1_3}
     &=
      e^{-2(β_\circ^3-β_\circ^1)}\bigg(\p{_\circ^3_1}
        -\frac{\p{_\circ^2_1}\p{_\circ^3_2}}{2p_\circ^3-2p_\circ^2}\bigg)\frac{1}{2p_\circ^3
        -2p_\circ^1}
       e^{-τ(2p_\circ^3-2p_\circ^1 )}
     \nn
\\
     &\hspace{1em}
      +O(e^{-τ(2p_\circ^3-2p_\circ^1 +ν)})
     \,,
\label{16VI15.6}
\end{align}
where the functions $\{P\indices{_\circ^i_a}\}_{1\le a < i\le 3}$ depend only on $x^i$.

The remaining functions $p_\circ^1$, $β^1_{\circ}$,  $\p{_\circ^3_2 }$ and $\p{_\circ^3_1 }$ are then determined from the asymptotic constraint equations:
\begin{align}
 \label{final_my_p_cond}
    p_\circ^1 &=-\frac{p_\circ^2 p_\circ^3}{p_\circ^2+p_\circ^3}\,,
\\
β^1_{\circ,3}
 &=-(p_\circ^2+p_\circ^3)^{-1}(p_{\circ,3}^2+p_{\circ,3}^1+β_{\circ,3}^2(p_\circ^1
 +p_\circ^3)+β^3_{\circ,3}(p_\circ^1+p_\circ^2))\,,
 \\
 \p{_\circ^3_2_{,3}}&=2\left(G_{2c}p_{\circ,2}^c+β_{\circ,2}^d p_\circ^f G_{df}\right)\,,
\\
 \label{28X14.1}
\p{_\circ^3_1_{,3}}&=-\p{_\circ^2_1_{,2}}+2\left(G_{1c}p_{\circ,1}^c+β^d_{\circ,1}p_\circ^f G_{df}\right)
 \,.
\end{align}
Here $G_{ab}$ denotes the constant matrix $G_{ab}=-\sum_{c\neq d}δ^c_aδ^d_b$.

Finally the Kretschmann scalar behaves as
\[
 R_{αβγδ}R^{αβγδ}=\left(\frac{16e^{4(β_\circ^1+β_\circ^2+β_\circ^3)}\big(p_\circ^2 p_\circ^3\big)^2  }{(p_\circ^2+p_\circ^3)^2}
       \big((p_\circ^2)^2+p_\circ^2p_\circ^3+(p_\circ^3)^2\big)+O(e^{-\nu\tau})
 \right)e^{\tau 4(p_\circ^1+p_\circ^2+p_\circ^3)}
 \,,
\]
and therefore, since $p_\circ^2p_\circ^3>0$, the curvature diverges as $τ\to \infty$. Along curves $γ(τ)=(τ,γ^i(τ))$, $τ\in [τ_0,\infty)$, fulfilling $|γ'^i(τ)|=O(e^{(p_\circ^i(γ^j(τ))-ε)τ})$ for some $ε>0$ and for $i=1,2,3$, the curvature diverges in finite proper time / length. 
\end{theorem}

\section{Conclusion}
We have constructed a large class of vacuum spacetimes containing a timelike singularity. The solutions asymptotically approach a timelike Kasner metric at each point $(x^i)$, which can be interpreted as the field of an infinitely extended thin rod, with positive mass for $J\neq1$ and negative mass for $J=1$  \cite{Israel1977}. As the Kasner exponents now depend upon the coordinates $x^i$ the solutions might represent the field of more complicated, non-symmetric and non-static, one-dimensional sources.

We have also noted that the cosmological billiards arguments of Damour, Henneaux, and Nicolai \cite{Damour2003} are not directly applicable to this case, because of the transformation of asymptotically infinite potential walls into infinite wells. 
One should keep in mind the results of Parnovsky~\cite{Parnovsky1980,Parnovsky1999}, who applied the original procedure used by BKL to the timelike case, and concluded that the heuristic construction of chaotic singularities remains applicable. It would be of interest to resolve this apparent contradiction.

In \cite{Shaghoulian2016} the authors argue, using a model Bianchi IX spacetime, that the change of sign of some of the wall terms is an artifact of the Iwasawa decomposition and that the affected walls vanish in a different gauge. It is not clear to us whether their arguments apply to the general inhomogeneous case.

\bigskip

\noindent{\sc Acknowledgements:} Supported in part by a uni:docs grant of the University of Vienna.
We are grateful to W.~Piechocki for drawing our attention to~\cite{Parnovsky1980,Parnovsky1999} and to P. Chru\'{s}ciel for helpful comments and discussions.
\printbibliography
\end{document}